# Ablation de ZnO par laser UV (193 nm) : Nano-agrégats en phase gazeuse.

I. Ozerov[1], A.V. Bulgakov[2], D. Nelson[3], R. Castell[4], M. Sentis[5] et W. Marine[1]

[1] *GPEC - UMR 6631 CNRS, Faculté des Sciences de Luminy, Case 901, 13288 Marseille Cedex 9, France*
[2] *Institut de Thermophysique, Novossibirsk, Russie*
[3] *A. F. Ioffe Institut, 194021 St.-Pétersbourg, Russie*
[4] *Université Simon Bolivar, Caracas 1080-A, Venezuela*
[5] *LP3 - FRE 2165 CNRS, Faculté des Sciences de Luminy, Case 917, 13288 Marseille Cedex 9, France*

**Résumé.** La condensation de nano-agrégats d'oxyde de zinc en phase gazeuse est mise en évidence lors de l'ablation de ZnO massif par laser ArF pulsé. Nous comparons l'évolution spatio-temporelle de la forme du panache d'ablation (plume) de ZnO sous vide et sous atmosphère de gaz de couverture (oxygène et/ou hélium) à partir des images CCD et des résultats issus d'analyses spectroscopiques. L'expansion du plasma et la croissance des nano-clusters sont influencées par l'effet du confinement de la plume dû aux collisions entre les particules ablatées et les molécules de gaz ambiant ainsi que par les réactions chimiques dans le cas de l'oxygène. Le spectre de rayonnement du plasma est constitué principalement par l'émission d'atomes excités de Zn neutre. Nous avons observé la photoluminescence des nano-agrégats en suspension dans le gaz ainsi que leur décomposition par laser ArF.

## 1. INTRODUCTION

L'oxyde de zinc est un semi-conducteur à large gap direct (3,3 eV) ayant une grande énergie de liaison d'exciton (60 meV). Les propriétés optiques de ce matériau le rendent très prometteur pour des applications dans le domaine de l'optoélectronique. Récemment, un effet laser très efficace à température ambiante a été obtenu à partir de nano-agrégats de ZnO organisés dans des couches minces [1]. Le contrôle de la composition, de la structure et des tailles des nano-agrégats est nécessaire pour le développement de lasers et de dispositifs optoélectroniques.

L'ablation laser en présence d'un gaz de couverture est une méthode très efficace de dépôt de films minces de différents matériaux nanostructurés [2,3]. Les films de ZnO préparés par cette méthode montrent une excellente qualité optique [4]. Lors de la synthèse de films de nano-agrégats, les processus de condensation et de cristallisation de ces derniers ont lieu dans un gaz ambiant (oxygène et/ou hélium) et les nanocristaux ainsi formés en phase gazeuse arrivent déjà refroidis sur le substrat [3,4]. Les propriétés des clusters obtenus peuvent être contrôlées par les paramètres du laser d'ablation (fluence, durée d'impulsion, longueur d'onde) et par les conditions du gaz ambiant (pression, nature, réactivité chimique, flux). Malgré les succès obtenus dans le dépôt de films nanostructurés de ZnO par ablation laser [4,5], l'expansion spatio-temporelle de la plume sous gaz de couverture, la condensation des nanoclusters et leurs propriétés n'ont pas été suffisamment étudiées. Les bandes de photoluminescence attribuées aux particules d'oxyde de zinc lors de l'ablation dans l'hélium ont été observées [6], mais l'étude des propriétés physiques des nanoclusters en suspension dans l'oxygène et dans le mélange $O_2$-He n'a pas été rapportée. L'importance du processus d'expansion de la plume d'ablation et sa relation avec la synthèse des nanoclusters en phase gazeuse montrent la nécessité de poursuivre une telle étude pour maîtriser le dépôt des films.

## 2. DISPOSITIF EXPERIMENTAL

Une cible de ZnO (céramique) est placée sur un support rotatif à l'intérieur d'une enceinte à vide en acier inoxydable évacuée à l'aide d'une pompe turbomoléculaire. Après obtention d'un vide de l'ordre de $10^{-7}$ mbar, on réalise une circulation permanente du gaz ambiant (oxygène, hélium ou mélange des deux) dans l'enceinte à une pression constante fixée entre 0.5 et 50 mbar (pompage différentiel). Ce procédé permet de réduire l'accumulation des impuretés gazeuses et des nano-agrégats dans la chambre. L'ablation est réalisée à l'aide d'un laser impulsionnel ArF* (Lambda Physik LPX205 ; $\lambda$ = 193nm ; $\tau_p$ = 15ns), sous un angle d'incidence de 45° par rapport à la normale de la cible, sur une aire de 0,7 mm² avec la partie la plus homogène du faisceau sélectionnée par des diaphragmes.

Les images de la plume sont prises par une caméra rapide CCD intensifiée. L'émission du plasma est analysée par spectroscopie résolue dans l'espace et en temps. L'image de la plume laser est transférée par une lentille sur la fente d'entrée d'un monochromateur couplé à une camera CCD (Andor). Les signaux de photoluminescence et de décomposition des nanoclusters en suspension dans le gaz sont enregistrés avec le même système d'acquisition que l'émission de la plume.

## 3. EXPANSION DE LA PLUME D'ABLATION SOUS VIDE ET DANS UN GAZ AMBIANT

Lors de l'ablation laser, un plasma dense d'atomes, d'ions et d'agrégats formés de quelques atomes se développe sous forme d'une plume perpendiculairement à la surface de la cible. . A partir des résultats de la spectroscopie optique de la plume, les atomes neutres sont les particules les plus nombreuses et on peut négliger l'émission optique des ions et des nano-agrégats pendant les études de la forme de la plume. L'évolution typique de la plume est représentée sur la Fig. 1. Des séries d'images sont prises par caméra iCCD rapide à partir de la plume obtenue par irradiation d'une cible de ZnO massif à une fluence laser de 3,5 J/cm² sous vide (Fig. 1a), dans 5,5 mbar d'hélium (Fig. 1b), et dans 4 mbar d'oxygène (Fig. 1c).

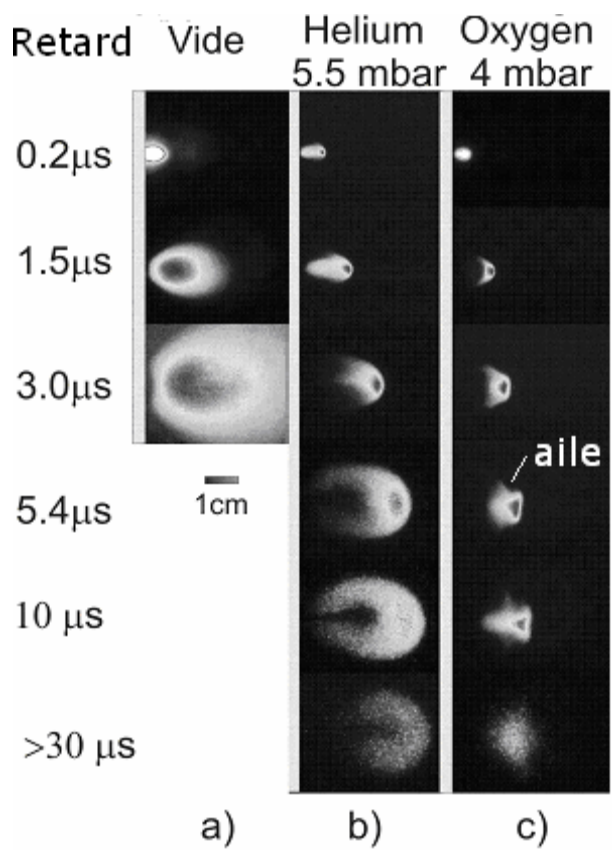

**Figure 1.** Images de l'expansion de la plume prises par caméra rapide iCCD à divers instants par rapport au début de l'impulsion laser d'ablation : a) Vide 2 $10^{-7}$ mbar ; b) Pression d'hélium 5,5 mbar ; c) Pression d'oxygène 4 mbar. (Fluence laser F = 3,5 J/cm² ). La cible est située à bord gauche des photographies.

L'évolution spatio-temporelle de la plume sous vide présente l'expansion libre avec une vitesse constante de propagation du centre de masse de la partie lumineuse ~$10^6$ cm/s (Fig. 1a). La plume s'élargit en trois dimensions avec des vitesses d'expansion axiale et radiale très proches. A cause de cette expansion on ne peut pas garder la même échelle d'observation pour des retards plus grands que 3µs. Par contre, l'expansion après l'ablation dans le gaz est plutôt axiale et la vitesse d'expansion diminue pour des retards plus grands que 1,5 µs. Ce processus de confinement de la plume est dû aux collisions des particules ablatées avec les molécules du gaz ambiant. L'effet de confinement de la plume favorise la condensation des nano-agrégats en phase gazeuse. Pour des retards plus grands que 5,4 µs la vitesse d'expansion de la plume sature. La plume dans hélium lentement continue son expansion (Fig 1b). Les particules de la partie frontale de la plume sont excitées par collisions avec les molécules du gaz ambiant et forment la partie lumineuse de l'arc de la plume. L'expansion dans l'oxygène est différente par rapport à l'hélium (Fig. 2c). A partir d'un retard de 5,4 µs la partie centrale de la plume s'arrête et on peut observer une formation d'instabilités en forme d'« ailes » en périphérie de la plume qui se développent en amont. Ce flux de particules en périphérie est dû à un gradient de pression. Nous expliquons l'apparition de ces instabilités par les réactions chimiques entre les particules de la plume et l'oxygène atomique créé lors des collisions. L'émission de la plume peut être observée jusqu'à des retards de 50 µs.

## 4. NANO-AGREGATS EN PHASE GAZEUSE

La présence d'agrégats de ZnO en suspension dans le gaz a été confirmée par l'observation d'une luminescence intense dans le trajet du faisceau laser ArF incident (fig. 2). La photographie CCD prise au moment du tir laser est présentée sur la fig. 2a. La cible (à droite en bas de l'image) est irradiée par laser ArF avec une incidence de 45°. Dans le trajet du faisceau laser (de la gauche en haut vers la droite en bas de l'image) on peut observer une émission dans la région spectrale de l'UV. Les spectres de cette émission dépendent fortement de la puissance du laser d'excitation. Pour une puissance faible, le spectre (fig. 2b) est constitué de deux bandes de luminescence centrées à 3,27 eV et à 3,06 eV avec des largeurs à mi-hauteurs de 0,17 eV et 0,18 eV respectivement. Ces bandes de luminescence correspondent aux transitions interbandes de ZnO massif sous excitation par laser pulsé [7]. Nous attribuons cette luminescence aux nano-agrégats de ZnO suffisamment refroidis et déjà cristallisés en suspension dans le gaz. La luminescence à 3,27 eV est issue à partir de la recombinaison des excitons libres dans des clusters de grande taille. La bande centrée à 3,06 eV correspond à des recombinaisons interbandes dans un matériau avec une forte concentration des porteurs (électrons et trous). La concentration élevée des porteurs dans un semi-conducteur provoque un effet de renormalisation de gap qui se manifeste par un décalage vers le rouge de la position spectrale d'émission interbande [7]. La présence à la fois de deux bandes peut être expliquée par la dispersion des tailles des nano-agrégats.

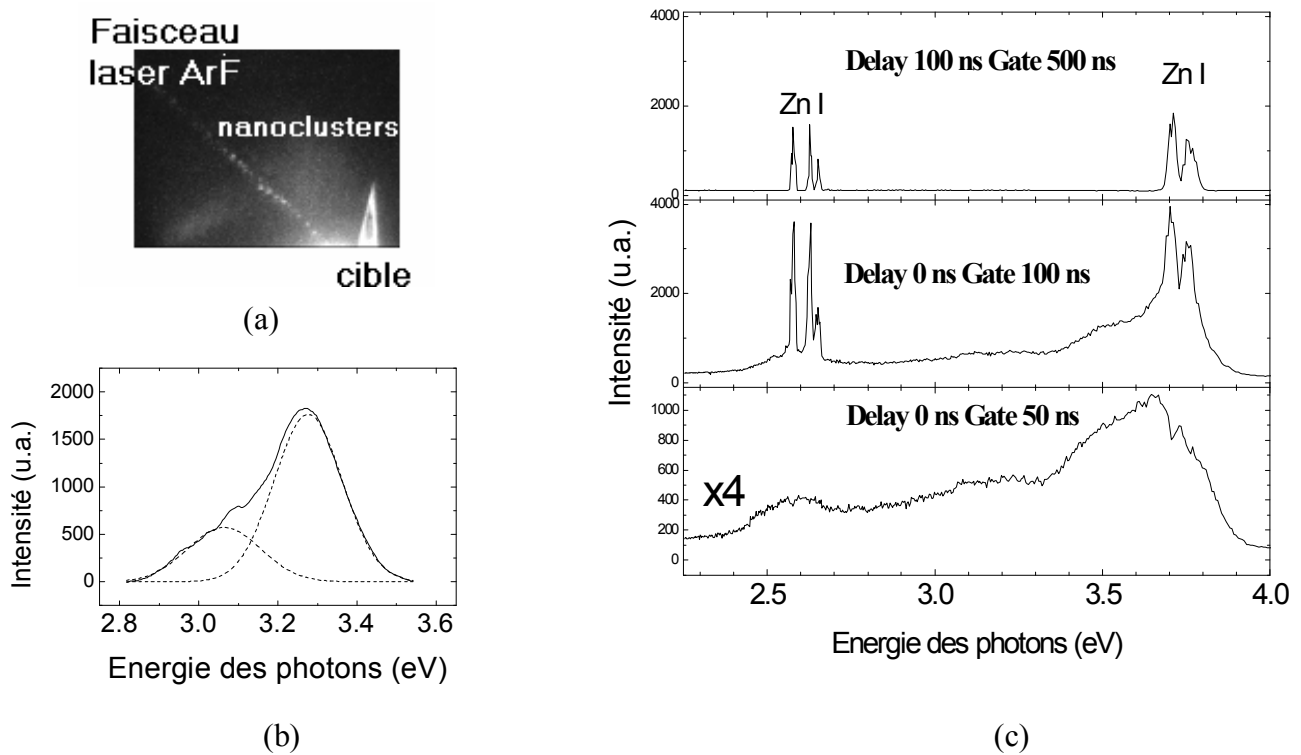

**Figure** 2. Photoluminescence et décomposition des nano-agrégats de ZnO en suspension dans l'oxygène (pression 4 mbar) induites par laser ArF : a) Photographie CCD des nano-agrégats de ZnO excités par l'impulsion laser incidente ; b) Spectre de photoluminescence des nanoclusters de ZnO (faible puissance de laser d'excitation) ; c) Spectres de photoluminescence des nanoclusters avec une forte puissance de laser d'excitation immédiatement après l'excitation (retard 0 ns) et leur décomposition aux instants suivants. Les raies d'émission d'atomes de Zn sont indiquées.

L'augmentation de l'intensité du laser d'excitation provoque un changement radical des spectres (fig. 2c). Pendant les premières 50 nanosecondes après l'impulsion laser, le spectre est non structuré et correspond à superposition de la luminescence des nano-agrégats avec un continuum dans une région spectrale comprise entre 2,3 et 3,8 eV. Ces spectres de continuum correspondent à l'émission observée pendant les stades initiaux de formation du plasma d'ablation (processus de Bremsstrahlung) [8]. Des raies fines d'émission des atomes neutres de zinc apparaissent 50 ns après l'impulsion du laser d'excitation. Pour des retards supérieurs à 100 ns, le spectre présente seulement des raies atomiques de zinc. Ces résultats mettent en évidence la décomposition des nanoclusters sous irradiation par laser ArF. Les atomes de Zn excités sont le résultat de leur évaporation à partir des nano-agrégats.

## 5. CONCLUSION

Nous avons utilisé l'ablation laser pour réaliser la synthèse de nano-agrégats d'oxyde de zinc. Les études de l'évolution spatio-temporelle de la forme de la plume d'ablation de ZnO montrent que les gaz ambiants influencent l'expansion de la plume. La présence de gaz ambiant (He, $O_2$) provoque un effet de confinement du panache d'ablation dû aux collisions entre les particules ablatées et les molécules du gaz ambiant. Lors de l'ablation en présence d'oxygène, nous avons observé des instabilités en forme d'ailes dans les parties périphériques de la plume. Ces instabilités dans le plasma sont expliquées par les réactions chimiques entre les particules ablatées et l'oxygène atomique créé dans les collisions. Le confinement de la plume par le gaz ambiant favorise la condensation des nano-agrégats. La condensation et la cristallisation des nanoclusters de ZnO en phase gazeuse sont mises en évidence. Par excitation laser ArF, la photoluminescence d'agrégats en suspension dans le gaz est observée sur le trajet du faisceau laser. Le spectre de cette luminescence correspond aux recombinaisons interbandes dans le ZnO nanocristallin. L'augmentation de l'intensité du laser d'excitation provoque l'apparition de raies d'émission des atomes de zinc. Ces atomes excités sont issus de la décomposition des nano-agrégats par le laser ArF.